\documentclass[12pt]{cernrep}
\usepackage[dvips]{epsfig}
\begin{document}
\title{LCG Monte-Carlo Data Base\\
       (LCG Generator Services Subproject)}
\author{
  P.~Bartalini$^1$
  L.~Dudko$^2$,
  A.~Kryukov$^2$,
  I.~Seluzhenkov$^3$
  A.~Sherstnev$^2$,
  A.~Vologdin$^2$
 }
\institute{
   $^1$CERN, Geneva, Switzerland,\\
   $^2$Moscow State University (MSU), Moscow, Russia,\\
   $^3$Institute for Theoretical and Experimental Physics (ITEP),
   Moscow, Russia.
 }
\maketitle

\begin{abstract}
We present the Monte-Carlo events Data Base (MCDB) project and its 
development plans. MCDB facilitates communication between authors of 
Monte-Carlo generators and experimental users. It also provides a convenient
book-keeping and an easy access to generator level samples.
The first release of MCDB is now operational for the CMS collaboration. 
In this paper we review the main ideas behind MCDB and discuss 
future plans to develop this Data Base further within the CERN LCG framework. 
\end{abstract}

\section{Problem description}

One of the most general problems for the experimental high energy physics 
community is Monte-Carlo (MC) simulation of physics processes. There are 
numerous publicly available MC generators. However, the correct MC simulation 
of complicated processes requires in general rather sophisticated expertise 
on the user side. Often, a physics group in an experimental 
collaboration requests experts and/or authors of MC generators to create MC 
samples for a particular process. Furthermore, it is common that the same 
physics process is investigated by various physics groups neeeding the same 
MC event samples. The main motivation behind the Monte-Carlo Data Base (MCDB) 
project is to make MC event samples, as prepared by experts, available for 
various physics groups.

There are a number of useful aspects that motivate setting up a central MC Database.

\begin{enumerate}

\item 
Correct and reliable MC event generation of the most processes of 
interest requires significant expertise. Moreover, most MC generators, 
in particular those calculating higher order perturbative corrections, 
require a significant amount of computer and human resources. By means of 
MCDB, samples prepared by experts can be distributed easily and used 
as many times as needed.

\item 
Public availability of common event files helps to speed up the
validation procedure of the events.

\item 
A central and public location where well-documented MC events can be found 
would be very useful. It would also allow rapid communication between 
authors of MC events and their users.

\item 
The same MC samples for Standard Model processes can be used for multiple 
purposes, e.g. to study backgrounds to various new physics processes.

\item 
Files containing detector and beam-related backgrounds can also be kept in 
a common location.
\end{enumerate}

Historically, the first MCDB -- PEVLIB~\cite{pevlib} was established 
at CERN on the AFS file system. This database provided CompHEP~\cite{comphep} 
parton level events for CMS, but was lacking of a special interface for 
users. It was rather built as a set of directories where event samples were 
stored. Documentation for the samples consisted of ASCII files (README)
located in the same directories as the event files.

The next version of MCDB was established at Fermilab. This 
database was split in two independent parts:
\begin{itemize}

\item 
MC event files, stored via the FNAL tape system ENSTORE~\cite{enstore}.

\item 
Documentation for the events, publicly available via the 
Web~\cite{fnal_mcdb}.

\end{itemize}

The latest version of MCDB~\cite{lesh_mcdb} was developed 
and deployed in the CMS collaboration -- CMS MCDB~\cite{cms_mcdb}. 
This database includes web interfaces both for event files (enabling 
download and upload) and documentation. Its main goal is to store events, 
only at the parton level, generated by MC experts, to be used by the LHC 
community. Note that all the files from PEVLIB have been moved to CMS MCDB.

\noindent However, we can identify several potential weakness in CMS MCDB that 
motivate further developments of this project.
\begin{itemize}

\item 
CMS MCDB was designed to store parton level events. This implies that 
the size of event files should not be too large (typically smaller then 
100 Mb).

\item 
CMS MCDB does not support a SQL engine. Therefore this database can 
process keyword phonetic queries only. Complex search is not possible.

\item 
Support for few important processes is guaranteed at the moment, however the LHC experiments need to use further processes (a multiplicity of several hundreds is estimated for LCG MCDB). 

\end{itemize}
These aspects do not limit CMS MCDB at present. However, we expect that in 
a few years users will request a more powerful MCDB where these restrictions 
will be removed.

The next version of MCDB (to be used by all the LHC collaborations) is now under development in the context of the LCG Generator subproject~\cite{lcg_generator}, an Application Area activity developed in the framework of the simulation project. 
The new MCDB has to provide persistent 
storage of event samples with convenient public interfaces 
for users from the LHC community and experts or authors of MC generators. 
The main requirements for the new version of MCDB are the following:

\begin{enumerate}

\item 
LCG MCDB is based on a SQL database, therefore it is possible to 
keep deeply structured information and to treat sophisticated 
search queries.

\item 
The database should store both events with partons in the final state 
(partonic events) and events after hadronization and all decays (particle 
events). It also can keep some other types of events.

\item 
Users have access to the MC event samples and descriptions of the samples 
via Web interfaces.

\item 
Users may publicly discuss MC samples.

\item 
Formal validation of MC samples by experts.

\item 
An authorized user (author) may add, modify, or delete any information about 
his(her) own event samples or the samples themselves dynamically by a simple 
Web interface.

\item 
Application software should have a programming access to the samples.
     
\item 
New methods of event files uploading (in addition to the existing methods 
in CMS MCDB) should be implemented. These methods would be based on 
GRID technologies.
\end{enumerate}

%---------------------------------------------------------------------------------------------
\section{General conception and terms}
LCG MCDB uses ideas and experience of CMS MCDB~\cite{lesh_mcdb}. 
We will use the following notations in the note (they are marked out by 
bold).

\noindent{\bfseries Event File (sample)} - a file containing particle or
partonic events. These files consist the main contents of MCDB.

\noindent{\bfseries Article} - a document describing a set of samples. 
The main task of the article is to provide comprehensive information 
about event samples (connected with the article) supplied by an author. 
The article is written via the Web interface and is freely available 
via the Web on the MCDB Web site. 

\noindent{\bfseries MCDB License} - agreement between an author and 
the end-users about event samples available via the MCDB Web site.

{\vspace*{3mm}
In the LCG MCDB conception we subdivide all the information about events
into two types:

\noindent{\bfseries Event Meta-data} - information which describes all
events in a sample \emph{in general} (e.~g. beam description,
physics parameters, applied cuts). The meta-data store SQL tables
of MCDB.

\noindent{\bfseries Event Data} - the events themselves (the event data 
are stored into event files).

These types of the information do not have clear limits. For instance, 
number of particles in a partonic event sample can be considered as 
meta-information. But, after hadronization, the number of particle 
will change from one event to another and this parameter, certainly, 
does not belong to the meta-data. However, a particular sample always 
can be subdivided into meta-information and event information. 
The meta-data provides a basis to the SQL search in MCDB.

{\vspace*{3mm}
There are 5 different access ways to MCDB in the LCG MCDB conception.
Each way has specific rights and restrictions. We will describe these
access methods in terms of ``users``. These ``users'' can be both real 
persons and software modules.

\noindent{\bfseries End-User} - a user who works with MCDB (search, 
read, download) via Web interfaces. For the downloading of event 
samples the end-user should register on the MCDB Web site and accept 
MCDB License.

\noindent{\bfseries Author} - an authorized user, which can create and 
modify his/her own articles and upload new event samples. 

\noindent{\bfseries Moderator} - an authorized user which manage 
(add, remove, modify) authors profiles in MCDB. Moderators may
be appointed 

\noindent{\bfseries Administrator} - system administrator of MCDB.

\noindent{\bfseries Application software} - a part of the experimental
software, which has an access to event files in MCDB via special
API's.

%---------------------------------------------------------------------------------------------
\section{Main LCG MCDB components}
Main components of LCG MCDB are presented in the figure \ref{main}.
\begin{figure}
\begin{center}
\includegraphics[width=12.5cm]{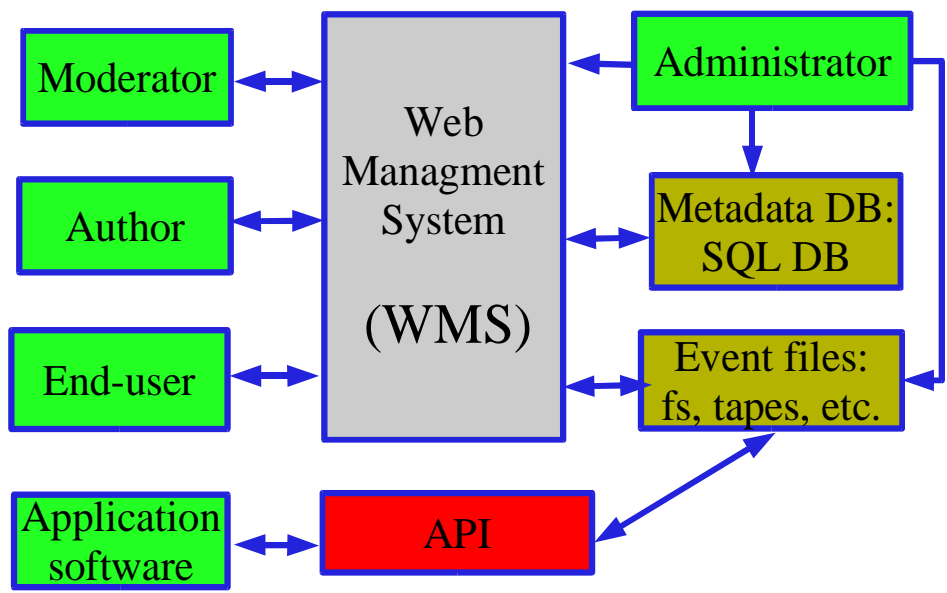}
\caption{The CERN $\bar{\rm p}$ complex.}\label{main}
\end{center}
\end{figure}

Event files are kept on a mass storage file system, a plain or 
a distributed file system. 

All meta-data are written to SQL tables of Event Meta-Data DB. This 
database will work under control of one of SQL DBMS (MySQL, PostgreSQL, 
or Oracle). 

WMS (Web Management System or Content Management System) provides
an access of end-users, moderators, and authors to read and write 
articles on the LCG MCDB Web site.

API (Application Programming Interface) is a library which provides a 
direct programming access of application software (for instance simulation 
software specific to a collaboration) to the event samples.

\section{MCDB interfaces}
\begin{figure}
\begin{center}
\includegraphics[width=10.5cm]{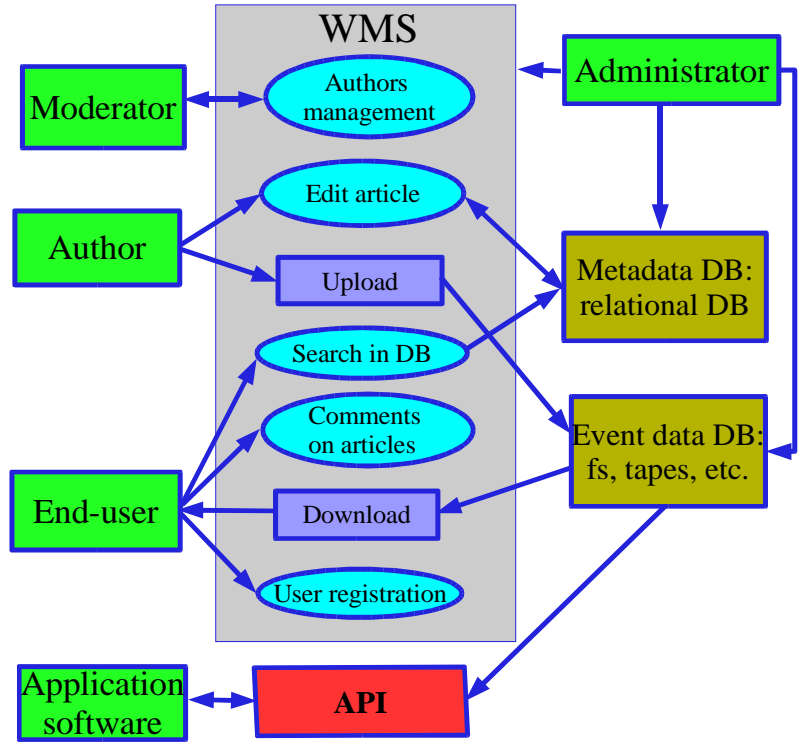}
\caption{MCDB interfaces to users and software}\label{interface}
\end{center}
\end{figure}

All interfaces of LCG MCDB are shown in the figure \ref{interface}.
End-users, authors and moderators work with MCDB via the Web
interfaces. All these interfaces are a part of WMS.

\noindent The end-user Web interface consists of several parts:
\begin{itemize}
\item 
Search form. 
Users mine data from LCG MCDB by filling a Web form. SQL requests 
based on the user requests are formed and sent to Event Meta-Data DB. 

\item 
Registration. 
A new end-user can register by filling out a special Web form. To 
successful register, an user should accept the MCDB License, 
which defines conditions of working with event samples in LCG MCDB. 
After that (s)he obtains a login name/password to access to the 
MCDB download area. 

\item 
Download samples. 
A registered end-user can download the samples to local machines. 

\item 
Comments on articles. 
A registered end-user can add comments on articles. A new comment is 
published on the Web. MCDB will have possibility to inform the article 
author automatically about the new comment by email. 

\item 
Web navigation. 
Articles are organized in a hierarchy tree structure on the MCDB Web 
site. The branches of the trees are defined to be the categories. 
End-user can navigate 
through this hierarchy. 
\end{itemize}

Each author has own profile in WMS and should pass an authorization 
mechanism in WMS, if (s)he would like to work in WMS. After authorization, 
the author can work with his/her own articles (create, modify, publish 
or interdit the access to articles in the LCG MCDB Web site). The author 
can upload new event files via the Web interface. The LCG MCDB Web 
site also has a special Web form to register new authors. 

The moderator Web interface allows to manage (create, remove, 
modify) the author's profiles and articles. 

Administrator supports WMS, Event Meta-Data DB, and event files. 

The application software has an access to event samples in LCG MCDB 
directly (for instance, by ftp or gridftp). The API organizes such 
connections. By default the API can download a full event sample. Application 
software can request a particular number of events. 
%If API can process data in the event file, API supplies a requested number of event. 

%---------------------------------------------------------------------------------------------
\subsection{WMS technical description}
This section describes the future internal structure of WMS. We plan to 
develop our own variant of WMS. %There are some reasons for that:
%\begin{itemize}
%\item
Although there are many WMS on the market, most of them 
are created for developing of Web portals (for example, for on-line 
stores or entertainment industry). No specific software fulfilling the 
only the MCDB requirements can be identified. Toolkits may have to be
customized.

e can not find a specific 
software with all (and only) functionality necessary for our goals. 
Therefore, we need to customize the toolkits (in many cases very deeply).

%\begin{itemize}
%\item
%Another's code is always a puzzle. If we plan to change different 
%blocks of a WMS, we have to know the code thoroughly. If the code is 
%large, it will be an apparent problem. 
%
%\item
%Versioning (and fast patching) also will become a problem. After new 
%releases of a used code we will have to customize the code repeatedly.
%\end{itemize}
Our originally developed code for WMS
will consist of separate blocks, which will 
perform a specific role in WMS. Certainly these separate 
blocks can be adopted from the existing software of the CMS 
MCDB or from other WMS (WebGUI, Metadot, etc.).

Now we mark out nine isolated and independent blocks 
which can be realized as a set of subroutines (Perl scripts)
with known input and output.
\begin{enumerate}
 
\item
Authorization.
This block sets permissions for each user according to the
user login and password. 
 
{\bf numb\_mode = general\_ACTH(login, password)}
  
where {\bf num\_mode} could be:\\ 
0 (administrator),\\
1 (moderator),\\
2 (author, with login name confirming the permission of the 
   particular author),\\
3 (end-user, after the ``License Agreement'').\\
As in CMS MCDB this block will be based on the CERN AFS passport
base. %???
 
\item 
Authors management.
This block manages the information about the MCDB authors and can 
consist of three requests (with interfaces via Web forms): 

{\bf 
author\_create(id,'login\_name','Last name','First Name','e-mail',...)\\
author\_modify(id,'new\_login\_name','new\_name','new\_e-mail',...)\\
author\_delete(id)
}
 
\item
Meta-data management. 
Authors should save meta-data about event files in SQL tables of LCG 
MCDB. After that post management block creates an article from the data.
Basically the meta-data management system has to allow to store, 
modify and remove the meta-data in the SQL database. Since the data 
are very closely related with articles we will give prefix ``article'' 
to the scripts in the block. These scripts will be realized as Web forms. 
In the future we plan to develop a procedure of automatic parsing of 
event files to fill some fields in the Web forms. Certainly, this 
operation will apply some requirements on format of the event files: 
 
{\bf
article\_create(article\_id,'category\_id1:category\_id2:...','author1:author2:...',...)\\
article\_modify(article\_id,'category\_id1:category\_id2:...','author1:author2:...',...)\\
article\_delete(article\_id)\\
article\_parser(article\_id)
} 
    
\item
Category management. 
This block manages categories and sub-categories. According 
to the goals of LCG MCDB, we intend to organize the categories
in several hierarchical levels

{\bf
category\_create(id,up\_category\_id,'category\_name') \\
category\_modify(id,up\_category\_id,'category\_name')\\
category\_delete(id,up\_category\_id,'category\_name')
 }

\item
Comments management. 
This set of subroutines creates and manages users comments 
to articles.
  
{\bf
comment\_create(article\_id,'comment\_body')\\
comment\_answer(article\_id,comment\_id,answer\_author\_id,'answer')\\
comment\_modify(article\_id,comment\_id,answer\_author\_id,'answer')\\
comment\_delete(article\_id,comment\_id)
} 
   
\item
Templates management block.
Each article requires a set of information about event sample. 
Articles in different physics categories can have a different form and 
the different minimal set of information. This block manages templates 
which make a pattern to articles in a particular category (or a set of 
categories) and defines a form how articles will be looked on the Web. 
There are two subparts in each template. The  first part defines the Web
form for the submission of an article to a particular category. The second 
part defines an HTML format for publication of articles from a particular 
category.
 
Functions to work with templates:\\
{\bf
template\_new(templ\_id,'category\_id1:category\_id2:...','fields')\\
template\_delete(templ\_id,'category\_id1:category\_id2:...')\\
template\_modify(templ\_id,'category\_id1:category\_id2:...','fields')\\
template\_html\_modify(templ\_id,'category\_id1:category\_id2:...','html\_cod')
}
 
\item
Post management.
An author submits and edits information about event samples to the SQL 
database by the articles management block. Finally, these information 
should be reorganized  to an article (statically or dynamically,  
depending on templates) and published on the LCG MCDB Web site. The main 
purpose of this block is to compile an article from information kept 
in Event Meta-Data SQL DB and post it to the LCG Web site. This block 
will allow to remove articles from the site if the author decides to do it.
 
{\bf
post\_article(article\_id,'category\_id1:category\_id2:...',post/delete)
}
 
\item
Uploading block.
This block manages the files attached to article. There are several 
methods to upload event samples by using different protocols (HTTP, ftp, 
ftpgrid). This block also will realize requests to (re)move files 
according to permissions and storage system optimization 
procedures.
 
\item
Log system. For security and debug reasons WMS needs to 
keep different levels of information about transactions
in LCG MCDB. The main function of this block is to collect
log information (to log files) from all other WMS blocks 
according to a debug level in a WMS configuration file. 
\end{enumerate}

\subsection{Choice of SQL DBMS for LCG MCDB}
After the comparison of different SQL databases we decided to 
apply MySQL~\cite{mysql} as SQL DBMS for LCG MCDB. MySQL offers 
many advantages for our project in comparison with other 
SQL database systems (Oracle, PostgreSQL, etc.):

\begin{itemize}

\item
MySQL is supported as standard software in the CERN LCG collaboration, 
in contrast to PostgreSQL.

\item
Open-source and free software: MySQL can be used, modified,
and distributed by everyone free of charge for any purpose,
be it private, commercial, or academic.

\item
Better support: in addition to strong support offerings, 
MySQL have a vibrant community of MySQL professionals and 
enthusiasts whose advice can be used during developing.

\item
Reliability and stability: unlike many proprietary databases, 
it is extremely common for MySQL users to report that MySQL 
has never crashed for them in several years of high 
activity operation. 

\item
Cross platform: MySQL is available for almost all Unix flavours and for Windows.

\item
GUI database design and administration tools: several high
quality GUI tools exist to both design and administer the database.

\end{itemize}

We plan to organize communication between WMS and SQL by 
means of separate library of PERL subroutines. Each block 
in WMS will correspond to a set of subroutines in this library
which will construct the SQL queries from WMS. The library of 
WMS-SQL interface for the LCG MCDB project will be described 
comprehensively in future technical documentation for the 
developers of the project.

%---------------------------------------------------------------------------------------------
\section{Unified event meta-data format}
The proposed conception of LCG MCDB demands no specific requirements 
on a format for the event files (it can be ntuples, plain text 
files, etc.), since this project does not intend to 
manipulate with events themselves. 

It would be highly useful to write the meta-data information in a unified format. 
This would allow to treat the event files 
automatically in LCG MCDB. For example, such format would allow to fill 
fields in SQL tables of Event Meta-Data DB automatically. It would 
simplify writing
of articles in LCG MCDB and decrease a probability of errors in MCDB 
and the articles. Unification of event file formats would allow to
develop a reliable and simple interface to the experimental environments.
%and eventually the format will simplify life of experimentalists
%in the LHC community too.

Initial requirements for the format:
\begin{itemize}

\item
Platform independence; 

\item
An uniform syntax for any types of meta-data stored in the event files.

\item
The syntax should be extensible (allow to add some new 
information to files with minimal changes in files). 

\item
Simple I/O and parsing by already existed and well-maintained 
software tools (XML parsers, etc.).

\end{itemize}
One of the variants for the format has been proposed in~\cite{hepml}.

The only requirements on meta-data contents in event file is 
``self-description''. The meta-data, kept in an  event file, 
should provide understanding what the events are and 
how users can use them.

%---------------------------------------------------------------------------------------------
\section{Meta-data and the query interface model}
If an end-user wants to take data from MCDB, at first, (s)he 
has to compose a query to Event Meta-Data SQL DB to get a list 
of articles with the requested parameters. There are some different 
types of interfaces to build this query. We plan to organize 
a combination of two methods in LCG MCDB. The articles in MCDB will 
be sorted out to a tree structure of categories. In the ideal 
situation each category is related with a class of physical 
processes (as it has been done in CMS MCDB). The end-user will 
browse the tree and will find the necessary class. The main task here 
is to choose interesting (for the end-user) category of the
processes. Articles in LCG MCDB will not be connected with the
categories (i.e. one article can fall into several categories).

When the end-user has chosen a category, (s)he may build a SQL 
query by simple ``language'' interface. This search query will 
be processed just in the frames of the category. 
%As we plan the main contains of LCG MCDB will be samples with
%partonic events and with events after showering/hadronization.
Since the information which describes these events is very heterogeneous, 
the proposed scheme allows to simplify syntax of the interface in 
the separate categories.

\noindent Preliminary list of information which will be available
to search in LCG MCDB:
\begin{enumerate}

\item
{\bfseries Initial state}: proton-proton, ion-ion, maybe
other beam information for some specific samples (machine
and beam related backgrounds).

\item
{\bfseries Type of the final state}: partonic level, parton
level plus showers, particle level (after hadronization).

\item
{\bfseries Generator specific parameters}: showering and 
hadronization models and their parameters, initial value of 
random generator in PYTHIA (or other MC generator), 
other parameters specific for a MC generator, etc.

\item
{\bfseries Physics which is related to the sample}: for 
instance, Higgs in MSSM, extra dimensions, gauge bosons
production, top quark physics, etc.

\item
{\bfseries Physics parameters for the process}: values of 
couplings, masses, widths, and other properties of particles 
(W- or Z-bosons, etc.), CKM matrix elements, and so on. 

\item
{\bfseries Formal information}: who, when, and how (by 
which generator) was created this sample, how many events 
in the sample, which format was been used, etc. 

\item
{\bfseries Applied cuts, the process cross section 
and cross section errors}.

\item
{\bfseries Direct links to event files}.

\end{enumerate}

The first two points will be encoded in the main tree of the LCG MCDB 
site. Other information will be collected for SQL queries by the 
``language'' interface.

Also authors may provide extra information in the samples
description using the special Web form (some author's free 
comments about the samples). This information will not be
used by the SQL queries of end-users for article searching.
It will only get to an article related with the sample as
an extra comment.

%---------------------------------------------------------------------------------------------
\section{How it works}
User interface to LCG MCDB will be organized as a Web site. To
search for some article on the Web site, a user browses through
the tree of (sub)categories (process classes) and fills out the 
search Web form. The result of the search will be a list of 
articles satisfied the queried conditions. After that (s)he can
read the selected articles. The article contains direct references
to event samples. If the end-user already registered (and 
accepted the MCDB License), (s)he can download the event
samples to a local machine and add comments to the articles. 
Unregistered users may register directly on the LCG MCDB Web site. 

Any author must register by filling out the author registration 
Web form. By the registration the author also accepts conditions 
of the MCDB License. The registration of new author should be 
approved by a moderator. Authorized author can work with his/her 
own articles:
\begin{itemize}

\item[--]
To create an article that describes the uploaded event file. First 
of all, the author uploads a new event sample. The article 
consists of two part. The first part is filled automatically 
by parsing event meta-data from the uploaded file. The second 
part is a comment supplied by the author. 
If the event sample is written in a not recognized 
format (by a WMS script), author fills all fields in 
the form by hands. 

\item[--]
To modify an existing article. The author chooses it from 
the list of his/her articles. The author can modify, add, or 
remove comments to the article and upload/remove event 
samples.

\item[--]
To make the articles accessible to the Web. The author publishes 
the articles in the LCG MCDB Web site. After that, the article will 
be available for the end-users. Also the author can disable any own 
article in this Web site (but it is still kept in MCDB 
itself and author can finish or modify it later and publish it 
again). 
\end{itemize}

Moderator manages author profiles via a special Web interface. 
Also (s)he can make disable or even remove articles. For example, 
if the sample or/and an article has bugs the moderator may disable 
the article temporarily. 

Application software have an access to the event data via API 
directly (e.g. by ftp anonymous session or by a GRID protocol 
in the future). Incoming data for the API are direct references 
to the event samples. These references can be obtained from the Web 
site. By default, API download full event data. if an event sample 
is kept in a standard and accepted by LCG MCDB format the 
application software can request a particular number of events.

%---------------------------------------------------------------------------------------------
\section{Milestones and requested resources}
The duration of proposal is two years. A plan of the first year
(by the quarter) is the following.

%00000000000000000000000000000000000000000000000000000000000000000000000000000000
\noindent First quarter: 
\begin{enumerate}
\item
Evaluation of software for LCG MCDB (WMS, SQL DBMS, etc.). 

\item
Design the LCG MCDB Web site.

\item
Design of structure and connections of SQL tables for LCG MCDB.

\end{enumerate}

%00000000000000000000000000000000000000000000000000000000000000000000000000000000
\noindent Second quarter:
\begin{enumerate}

\item
Development of Meta-data SQL DB (all necessary tables, 
subroutines to communicate with MySQL).

\item
Design and development of scripts for the meta-data
management and authorization blocks.

\item
Design and development of the author management, comments
management, and post management (in a reduced form) blocks.

\item
Deployment an customization of all necessary software
at the CERN. Deployment of the test version of LCG MCDB.

\end{enumerate}

The result of the first six months will be a test prototype of 
LCG MCDB. The software will be installed at the CERN and will 
include a first version of Event Meta-Data DB (all SQL tables) and 
all parts of WMS (all blocks will provide necessary options for 
testing only, full planned functionality will realize later on).

%00000000000000000000000000000000000000000000000000000000000000000000000000000000
\noindent Third quarter:
\begin{enumerate}
\item
Testing of the first LCG MCDB release. 

\item
Writing of LCG MCDB documentation, including HOWTO's for end-users 
and authors. 

\item
Development of a full version of the post management system. 

\item
Development of a full set of libraries. 
\end{enumerate}

%00000000000000000000000000000000000000000000000000000000000000000000000000000000
\noindent Forth quarter: 
\begin{enumerate}
\item
Development of the log system and templates management block.

\item
Development of tools for large file uploading  -- Uploading block. 
\end{enumerate}

%00000000000000000000000000000000000000000000000000000000000000000000000000000000
Tasks for the second year:
\begin{enumerate}

\item 
Creation of event Meta-data parsing mechanism. 

\item 
Development of API for application software.

\item 
Adaptation of LCG MCDB for LCG persistence system (POOL) and 
distributed storage systems.

\item 
Development of HEPML: specification (XML Schema) and parsing software 
including XSLT. 
\end{enumerate}

%00000000000000000000000000000000000000000000000000000000000000000000000000000000
\noindent {\bfseries Manpower}:
It is difficult to give precise estimation of manpower for the project.
Our estimation is 1.5-2 FTE per year. Nevertheless, we propose to start 
with 0.5 FTE from Russian side plus 0.5 FTE from CERN (1 FTE per year 
in total) with possible increasing of the manpower after six months.

\noindent {\bfseries Hardware and software requirements}:
Web server with large storage capacity (about 0.5-1.0 TB of local space) under Linux OS.
We need installed Apache and MySQL on the machine.

If the server will be under central administration, all MCDB developers 
have to included to /etc/sudoers with all necessary rights.

\noindent {\bfseries Periodicity of reports}:
Internal review should appear every six months.
\vskip 1cm

%---------------------------------------------------------------------------------------------

\end{document}